\documentclass[conference]{IEEEtran}
\IEEEoverridecommandlockouts
\usepackage{cite}
\usepackage{amsmath,amssymb,amsfonts}
\usepackage{algorithmic}
\usepackage{algorithm}
\usepackage{graphicx}
\usepackage{textcomp}
\usepackage{xcolor}
\usepackage{url}
\usepackage{mdwmath}
\usepackage{mdwtab}
\usepackage[all]{xy}
\usepackage{array}
\usepackage[tight,footnotesize]{subfigure}
\usepackage{stfloats}
\usepackage{bbm} 
\usepackage{adjustbox}

\def\BibTeX{{\rm B\kern-.05em{\sc i\kern-.025em b}\kern-.08em
    T\kern-.1667em\lower.7ex\hbox{E}\kern-.125emX}}
\begin{document}

\title{A Compact Low-Latency Systematic Successive Cancellation Polar Decoder for Visible Light Communication Systems}

\author{\IEEEauthorblockN{Duc-Phuc Nguyen\IEEEauthorrefmark{3}\IEEEauthorrefmark{2},
		Dinh-Dung Le\IEEEauthorrefmark{2},
		Thi-Hong Tran\IEEEauthorrefmark{2},
		Takashi Nakada\IEEEauthorrefmark{2},  
		Yasuhiko Nakashima\IEEEauthorrefmark{2}}
	\IEEEauthorblockA{\IEEEauthorrefmark{3}ETIS, UMR-8051, Université Paris Seine, Université de Cergy-Pontoise, France}
	\IEEEauthorblockA{\IEEEauthorrefmark{2}Graduate School of Information Science, Nara Institute of Science and Technology, Japan\\Email: (nguyen.duc-phuc@ensea.fr),(le.dung.ku9, hong, nakada, nakashim@is.naist.jp)}}

\maketitle

\begin{abstract}
	Channel polarization and Polar code are widely considered as major breakthroughs in coding theory because they have shown promising features for future wireless standards. The main drawbacks of Polar code are high-latency in decoding hardware, and unimpressive error-correction performance in case limited code-length is implemented. These two disadvantages limit implementation of Polar code in low-throughput wireless communication systems. In this paper, we propose a low-complexity low-latency hardware architecture for the soft-decision compact (16,11) Systematic Successive Cancellation Polar Decoder (S-SCD). Experimental results has shown that the latency of the proposed S-SCD improves 3.75 times and 2.75 times compared with conventional and 2b-SC architectures. Besides, it has also shown a better BER/FER performance compared with RS(15,11) code, which is applied widely in current VLC-based systems. 
\end{abstract}


\section{Introduction}
\label{intro}
Visible light communication (VLC) refers to short-range optical wireless communication using the visible light spectrum. VLC transmits data by intensity modulating optical sources, such as light emitting diodes (LEDs) and laser diodes, faster than the persistence of human eye \cite{rajagopal2012ieee,le2018joint,tuan2018demonstration}. LEDs are also increasingly being adopted in the general illumination market in both the commercial and residential segments, because of their advantages over competing lighting technologies in energy efficiency, longevity, color rendering capability, and environment factor \cite{jovicic2013visible}. However, VLC has certain shortcomings compared to traditional RF communication. The main drawback is that the achievable data rate drops sharply with increasing link distance, which limits the range of high data rate VLC use cases \cite{jovicic2013visible}. Fortunately, we can increase channel reliability and link distance by forward error correction (FEC) techniques \cite{le2018joint,le2018log,nguyen2018hardware}. 

In communication system, FEC is an error correction method by encoding data with redundant bits at transmitter. The redundant data enable receiver detect and correct some errors without asking the transmitter to re-transmit the data\cite{nguyen2018vlsi}. FEC techniques are also known as channel coding methods. Current optical networks employ FEC based on classical error-correcting codes such as Reed-Solomon (RS) or Bose-Chaudhuri-Hocquenghem (BCH) codes \cite{sakib2011optical}. Both RS and BCH codes currently use hard-decision-based receivers that have limited coding gain. Fig.1 shows block diagram of a typical low-data-rate VLC transmitter/receiver. In this system, a concatenation FEC solution is selected for the channel encode/decode. RS code is at Outer side, and at the Inner side is the Convolutional Code (CC). FEC solutions of different operation modes in VLC systems are well presented in \cite{rajagopal2012ieee}. 
Polar code is introduced as a low-complexity channel coding method that can achieve Shannon’s channel capacity for any binary-input symmetric discrete memoryless channel \cite{dizdar2016high}.
 Systematic polar code (SPC) were proposed by Arikan and are known for their improved bit error rate (BER) performance compared to the original non-systematic polar codes \cite{vangala2016efficient,arikan2011systematic,arikan2008channel,nguyen2018hardware}. The basic decoding algorithm for Polar codes is the Successive Cancellation (SC) algorithm, which is a non-iterative sequential algorithm with complexity $O(NlogN)$ for a code of length N. Due to low-complexity and high-performance, Polar code now is applied in many systems. The main drawback of Polar code is unimpressive error-correction performance in case of short code length is used. In this case, many approaches are introduced to enhance the performance of Polar code to make it feasible in systems which requires limited code lengths. 

In this paper, we propose applying Polar code as a FEC solution for VLC systems. From experimental results, we found that the Polar code outperforms RS code on error correction performance. We also propose a low-latency low-resource architecture for the (16,11) soft-decision Systematic Successive Cancellation Polar decoder (S-SCD). 

 \begin{figure}[h!]
	\centering
	\includegraphics[width=3.5in]{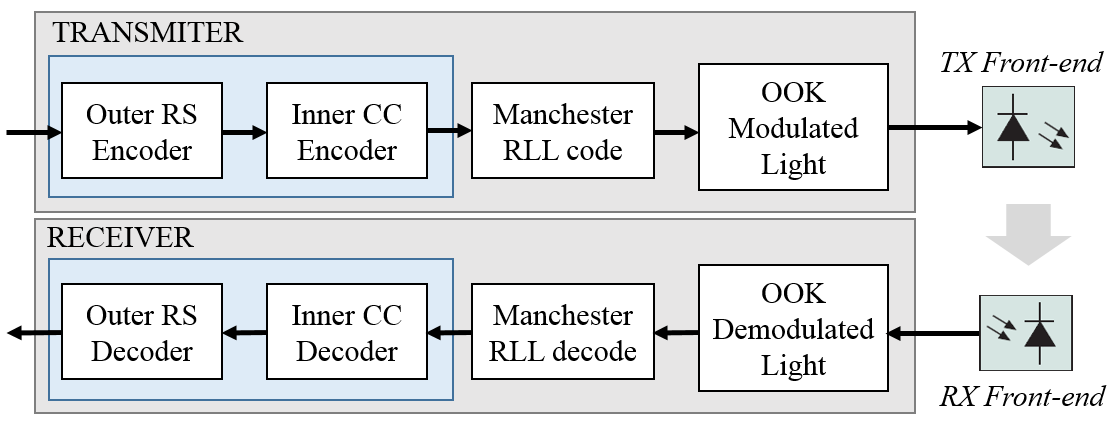}
	\caption{A low-data-rate VLC OOK transmitter/receiver}
	\label{fig1}
\end{figure}

\section{Polar Encoding/Decoding}

 \begin{figure}[h!]
	\centering
	\includegraphics[width=3.5in]{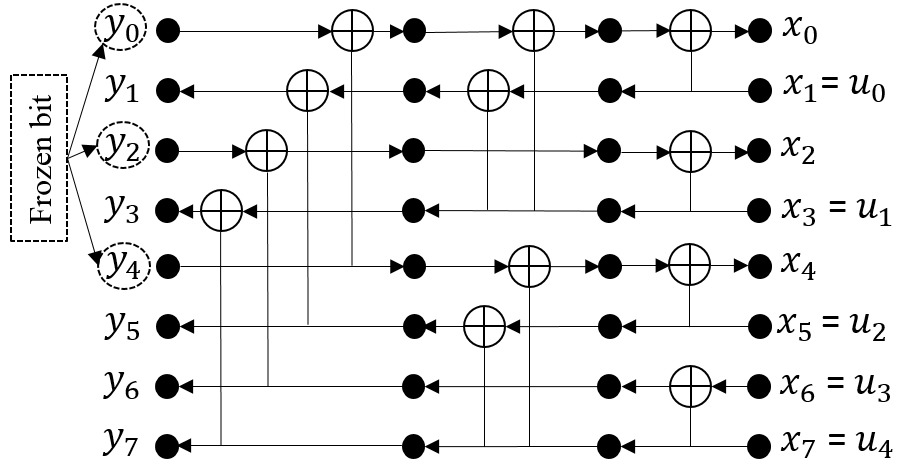}
	\caption{Systematic Polar Encoder (8,5) }
	\label{fig2}
\end{figure}

A polar code may be specified completely by $(N,K,F)$ where $N$ is the length of a code word in bits, $K$ is the number of information bits encoded per codeword, and $F$ is a set of indices known as information bit indices \cite{vangala2016efficient,vangala2015comparative}. For an $(N,K,F)$ polar code we describe below the encoding operation for a vector of information bits $u$ of length $K$. The rate of the code is $R=K/N$. Let $n = {\log _2}(N)$ and $G = {F^{ \otimes n}} = F \otimes \begin{array}{*{20}{c}}
.&.&.
\end{array} \otimes F$ (n copies) is the n-fold Kronecker product of Arikan’s \cite{arikan2011systematic,arikan2008channel} standard polarizing kerner, \[F = \left[ {\begin{array}{*{20}{c}}
	1&1\\
	0&1
	\end{array}} \right]\]
Then a codeword is generated as Equation \ref{eq01}.

\begin{equation}
\label{eq01}
x = u.G = d.{F^{ \otimes n}}
\end{equation}

Polar codes in their standard form are non-systematic codes \cite{arikan2008channel}. In other words, the information bits do not appear as part of the codeword transparently. A systematic polar code may be described as an equivalent to original polar code, except that the message vectors are mapped to codewords, such that the message bits are explicitly visible. Systematic Polar encoding of an information u of K bits, is the solution of Equation \ref{eq02}.

\begin{equation}
\label{eq02}
x = y.{F^{ \otimes n}}
\end{equation}

Where ${y_F}$ (message bit positions) and  ${x_{Fc}}$ (frozen bits positions) are the unknowns. There are exactly $N$ unknowns, shared between $x$ and $y$. In this paper, we implement a non-recursive systematic polar code which is introduced in \cite{vangala2016efficient}.

\begin{table}[]
	\centering
	\label{table_I}
	\caption{Processing Element (PE) Inputs and Output}
	\begin{tabular}{|c|c|c|c|c|}
		\hline
		SEL & $\hat{u}$ & La & Lb & Output (Out)            \\ \hline
		0   & x & x  & x  & $Sign (La.Lb).min({\lvert} La{\rvert}{\lvert} Lb{\rvert} $)\\ \hline
		1   & 0 & x  & x  & Lb + La                 \\ \hline
		1   & 1 & x  & x  & Lb - La                 \\ \hline
	\end{tabular}
\end{table}

Fig.\ref{fig2} shows an example of encoding of a (8,5) Systematic Polar Encoder (SPE). At the positions of frozen bits, the value of y is 0, and the data flow runs from left to right. At the position of information bit, data flow runs from right to left. Encoded bits are results at x. We can found that inside the encoded data $x$, it includes original information bits $u$. This encoder requires only \[(N/2).{\log _2}N\] XOR computation, which is the same as non-systematic Polar encoder. 

 \begin{figure}[h!]
	\centering
	\includegraphics[width=3.5in]{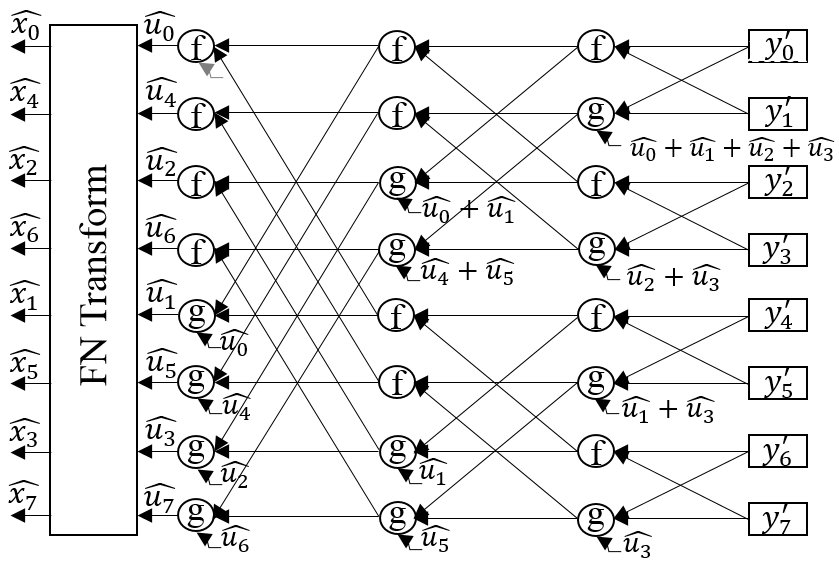}
	\caption{Systematic Polar Decoder (N=8)}
	\label{fig3}
\end{figure}

The most popular decoding algorithm for Polar code is the SC algorithm which is first introduced by Akiran \cite{arikan2008channel}. In fact, we employ the same SC decoder for both systematic and non-systematic codes. In both case, the decoder tooks input $y'$ and produces an estimate $\hat u$ of $u$. For non-systematic coding, the decoder stops after putting out ${\hat x_A}$. For systematic coding, decoder has an extra step of computing an estimate $\hat x = \hat u.G$ of $x$, and produced ${\hat x_A}$ as output. Fig.\ref{fig3} presents decoding diagram of a S-SCD.  
The S-SCD involves calculations using likelihood ratio (LR) values. The LRs are usually stored directly in floating-point variables. It is well-known to cause an underflow or an overflow. A popular solution to this problem is to store log-likelihood ratios instead of likelihood ratios. Real-valued calculations in log-domain is used for F function which is inside each processing elements (PE) (Fig.\ref{fig3}). Specifically, F function in log-domain is presented in Equation \ref{eq03}:

F function:

\begin{equation}
\label{eq03}
\begin{split}
F = add\_\log (x + y,0) - add\_\log (x,y)\\
\left\{ {\begin{array}{*{20}{c}}
	{z = y + \log (1 + \exp (x - y))\begin{array}{*{20}{c}}
		{}&{;x < y}
		\end{array}}\\
	{z = y + \log (1 + \exp (y - x))\begin{array}{*{20}{c}}
		{}&{;x \ge y}
		\end{array}}
	\end{array}} \right.
\end{split}
\end{equation}
  
Approximation form of F function: 
\begin{equation}
\label{eq04}
F = sign(x.y).\min (\left| x \right|,\left| y \right|)
\end{equation}

In hardware implementation, computing efforts based on logarithm and exponential are high-complexity. Normally, $F$ and $G$ functions can be implemented by simple logic gates and logic circuits by using approximation form as shown in Tab.\ref{table_I} and Eq.\ref{eq04}.

\section{Proposed Architecture}
\label{sectionIII}
\begin{figure}[h!]
	\centering
	\includegraphics[width=3.5in]{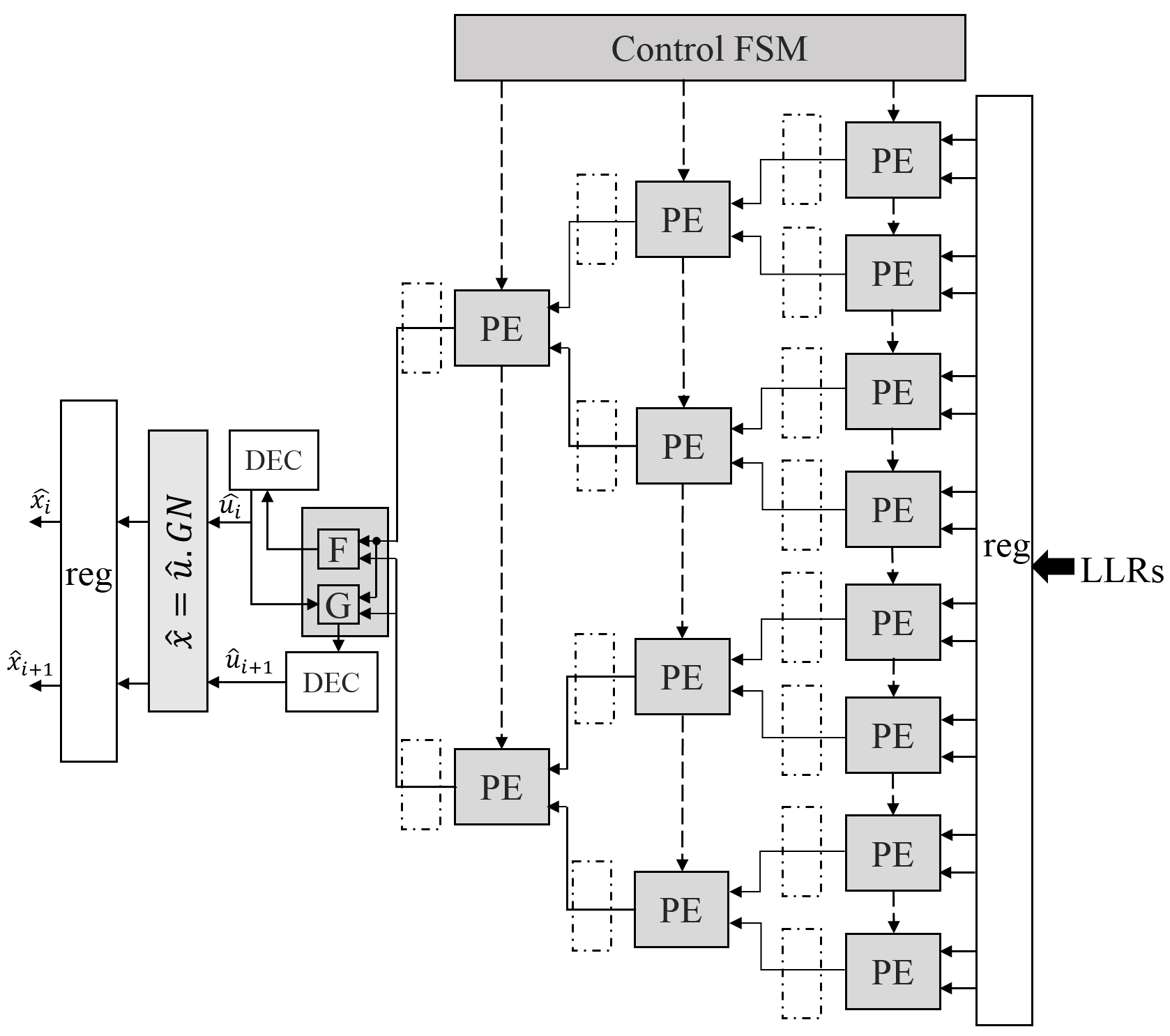}
	\caption{Proposed (16,11) S-SCD architecture}
	\label{fig4}
\end{figure}

\begin{figure}[h!]
	\centering
	\includegraphics[width=3.5in]{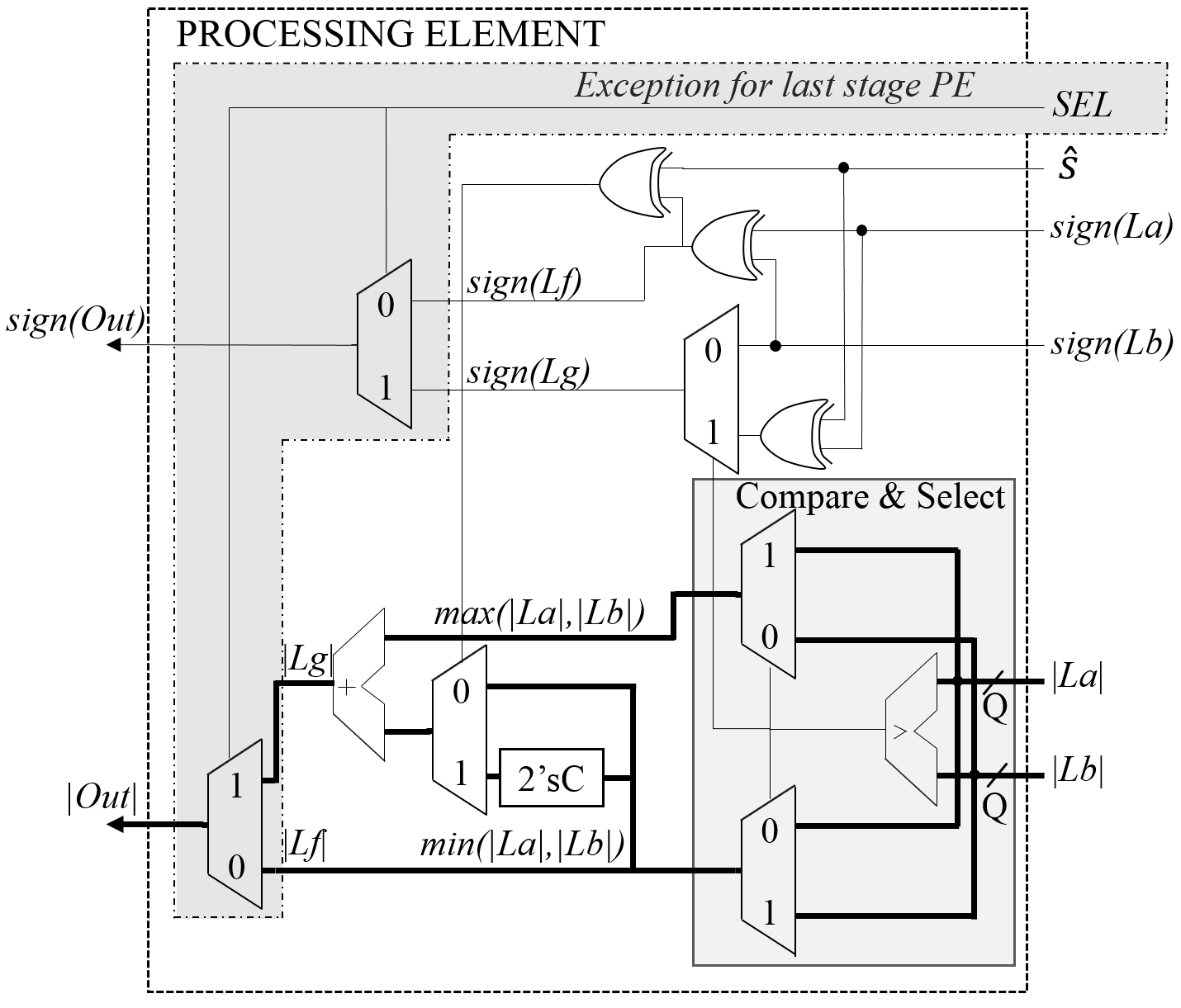}
	\caption{Processing element (PE) architecture}
	\label{fig5}
\end{figure}

The proposed architecture of (16,11) S-SCD is specified in Fig.\ref{fig4}. The architecture includes five main parts:
\begin{itemize}
	
	\item \textit{PEs (Processing Elements)}: Inside each $PE$ block, we implement one $F$ and one $G$ function. A $‘SEL’$ signal is created to select which output of either F-function or G-function circuit is the output of the $PE$. Fig.\ref{fig5} shows the specification of the $PE$.
	
	\item \textit{Modified $PE$}: At the last stage of $PE$ tree, we modify the architecture of a normal $PE$ by extract both outputs of F-function and G-function circuits as the outputs of $PE$. Decoded bit $u_i$ is forwarded to the input of G-function circuit to decode ${\hat u_{i + 1}}$
	
	\item \textit{Control Finite State Machine (FSM)}: This block implements a FSM to manage scheduling for whole S-SCD core. It assigns $SEL$ and $s$ signals, which control the operation of $PE$ tree network.
	
	\item \textit{Decoding (DEC) and FN transform}: $DEC$ and $FN$ transform logic circuit is implemented by simple combinational circuit. 
	
	\item \textit{Registers}: In proposed architecture, for each clock cycle, S-SCD finishes decoding 2 bits. For each event of positive clock edge, new input data will be loaded to $PEs$ of stage 3, and two decoded bits at the previous round will be stored in data registers.
	
\end{itemize}

\begin{figure*}[h!]
	\centering
	\includegraphics[width=5in]{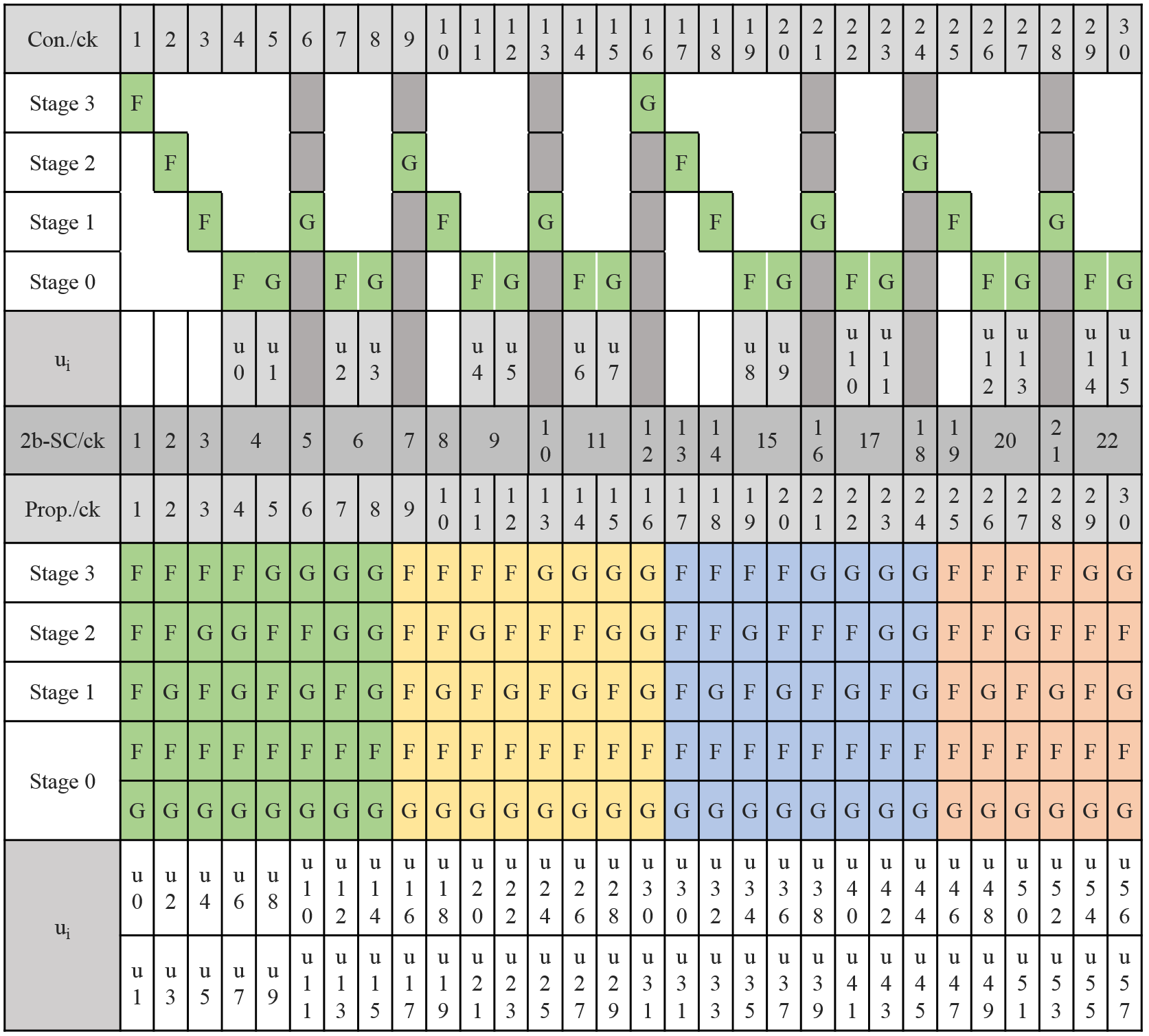}
	\caption{Proposed scheduling technique for the proposed (16,11) S-SCD}
	\label{fig6}
\end{figure*}
\begin{figure}[h!]
	\centering
	\includegraphics[width=3.5in]{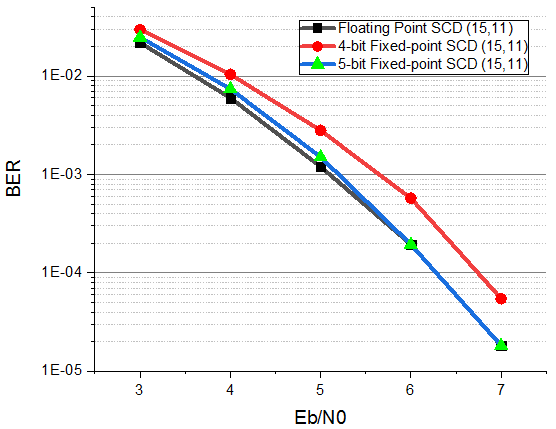}
	\caption{Performance of the fixed-point SCD at different numbers of quantization bit Q.}
	\label{fig7}
\end{figure}

\begin{figure}[h!]
	\centering
	\includegraphics[width=3.5in]{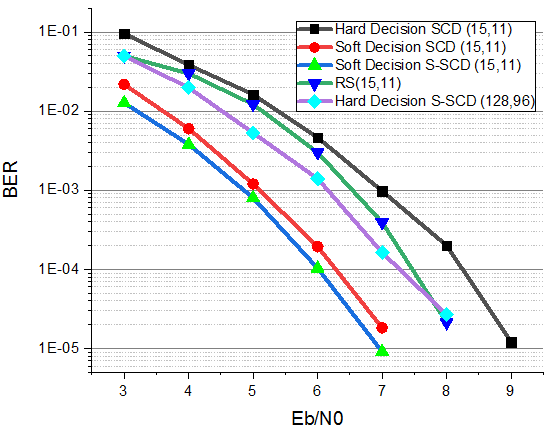}
	\caption{Bit-error-rate performance of the proposed soft decision S-SCD.}
	\label{fig8}
\end{figure}

\begin{figure}[h!]
	\centering
	\includegraphics[width=3.5in]{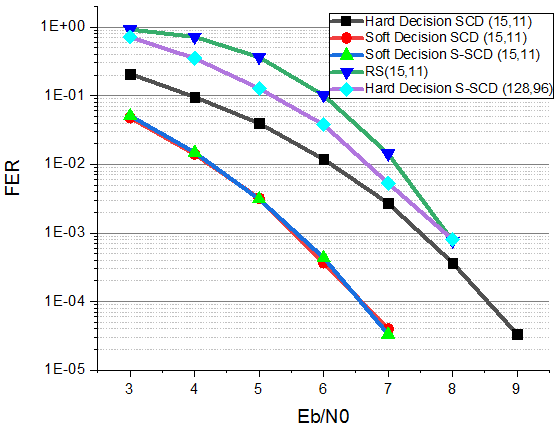}
	\caption{Frame-error-rate performance of the proposed soft decision S-SCD.}
	\label{fig9}
\end{figure}

Current hardware architectures for SCD focus on high-throughput \cite{dizdar2016high,leroux2011hardware,yuan2014low,nguyen2018vlsi,nguyen2018hardware} and they are expected to be applied in high-speed systems. On the other hand, VLC systems work mostly on low-data-rate (PHY-I) and medium-data-rate modes (PHY-II, PHY-III) \cite{rajagopal2012ieee}. Data-rate range varies from 11.67 Kb/s to 5 Mb/s with optical clock rate is set up to 7.5 Mhz. In this paper, we propose a low-latency, low-resource architecture for S-SCD. In this architecture, high-throughput is not the highest priority in design criteria. Specifically, we implement a compact (16,11) soft-decision S-SCD. The decoder shows a good performance with low-complexity of implementation. Code rate (N/K = 16/11) is also an equivalent code-rate with RS(15,11), which is the FEC solution in many modes of VLC systems. Fig.\ref{fig6} shows the processing scheduling technique of the proposed architecture. (16,11) S-SCD has four stages of PEs. In conventional architecture \cite{leroux2011hardware}, one clock cycle is dedicated for each PE stage’s processing, and two clock cycles are spent for the last stage. We propose all PEs of four stages to be processed within one clock cycles. For the last stage, we make some minor modifications to make the PE extracts two decoded bits at the output of the last stage’s PE.
\\

Furthermore, the proposed architecture is based on fixed-point calculations. Therefore, deciding number of quantization bit Q is very important. Fig.\ref{fig7} shows BER performance of 4-bit and 5-bit fixed-point (16,11) SCD compared with its floating point version. At Q=5, the fixed-point decoder shows a similar BER performance with the floating-point decoder.

\begin{table*}[h!]
	\centering
	\label{table_II}
	\caption{Latency of popular SCD architectures (In case of these architectures are applied in (16,11) S-SCD)}
	\begin{tabular}{|c|c|c|}
		\hline
		\textbf{Architecture} & \textbf{1st, 2nd bits decoding Scheduling} & \textbf{Number of clocks for decoding} \\ \hline
		Conventional \cite{leroux2011hardware}  & (F)-(F)-(F)-(F)-(G)                       & 30 clocks                              \\ \hline
		2b-SC \cite{yuan2014low}         & (F)-(F)-(F)-(F-G)                          & 22 clocks                              \\ \hline
		Proposed              & (F-F-F-F-G)                                & 8 clocks                               \\ \hline
	\end{tabular}
\end{table*}

\section{Experimental Results}

\begin{table*}[h!]
	\centering
	\label{table_III}
	\caption{Hardware synthesis results}
	\begin{tabular}{|c|c|c|c|c|}
		\hline
		& Logic Elements & Registers & Memory Bits & Fmax      \\ \hline
		10-bit S-SCD & 1108           & 181       & 0           & 58.11 Mhz \\ \hline
		6-bit S-SCD  & 700            & 117       & 0           & 66 Mhz    \\ \hline
		5-bit S-SCD  & 578            & 101       & 0           & 73.53 Mhz \\ \hline
	\end{tabular}
\end{table*}

Tab.\ref{table_II} summaries the number of latency clocks of the proposed S-SCD compared with conventional architecture and 2b-SC concept. The proposed (16,11) S-SCD requires only 8 clocks to finish decoding 16 bits; which reduces the latency 3.75 times and 2.75 times, compared with conventional and 2b-SC architectures respectively. However, maximum clock frequency of the proposed decoder is slower than if conventional and 2b-SC architectures are applied. This creates a trade-off between latency and clock frequency of the decoder.

However, as explained in Section \ref{sectionIII}, for a low-data-rate systems like VLC; latency seems to be put in higher priority compared with through-put \cite{nguyen2019fpga}. Fig.\ref{fig8} and Fig.\ref{fig9} show bit-error-rate (BER) performance and frame-error-rate (FER) performance of the proposed soft-decision S-SCD. These figures also show performance of the reference RS(15,11) and hard-decision SCD and S-SCD. Performance of soft-decision Polar decoder is much better than hard-decision decoder. Specifically, Fig.\ref{fig8} shows a 2dB better in coding gain between soft-decision and hard-decision (16,11) SCD, at BER=1E-4. Because RS(15,11) is widely used in VLC systems, we make a comparison between RS(15,11) which one symbol includes 8 bits; and hard-decision (128,96) S-SCD. In this case, not only BER, FER performances of hard-decision (128,96) S-SCD are better; the S-SCD also shows a better information utilization, in term of higher code-rate is used (128/96 compared with 15/11). In summary, we propose applying soft-decision (16,11) S-SCD, because of its low-complexity, and better BER/PER performance compared with current RS solutions in VLC systems.

Tab.\ref{table_III} summarizes results of hardware synthesis of the proposed (16,11) soft-decision S-SCD. With Q=5-bit, the proposed hardware can get maximum frequency around 73 Mhz while keeping low-resource, in which no memory bits are used. The synthesis results shown Tab.\ref{table_III} are achieved by synthesizing the proposed design with Quartus II software. The selected FPGA device is Altera Cyclone IV EP4CE115F29C7N.

\section{Conclusion}
In this paper, we have proposed a low-latency, low-resource architecture for the compact (16,11) soft-decision S-SCD. The proposed decoder has shown an improvement in latency compared with conventional and 2b-SC architectures. We have also shown that BER/FER performance of the proposed S-SCD is better than RS(15,11), which is current FEC approach in VLC systems. Moreover, hardware synthesis results have demonstrated that the proposed S-SCD is a low-complexity FEC solution. Therefore, the proposed decoder is quite suitable to be applied in VLC systems.
For near future works, we are building a full VLC system based on FPGA and customized VLC front-ends, in which the proposed S-SCD is also applied.

\section*{Competing interests}
The authors declare that they have no competing interests.
	
\section*{Acknowledgements}
This work was supported by JSPS KAKENHI Grant Number JP16K18105.

\bibliographystyle{IEEEtran}
\bibliography{IEEEexample}

\end{document}